\documentclass{llncs}
\usepackage[latin1]{inputenc}
\usepackage{amsfonts,amsmath,graphicx}
\usepackage{pstricks,pst-node}
\usepackage[expansion=false]{microtype}

\SpecialCoor

\psset{unit=9mm}

\definecolor{grid}{rgb}{0.8,0.8,1.0}
\definecolor{subgrid}{rgb}{0.9,0.9,1.0}

\psset{%
    gridcolor=grid,
    subgridcolor=subgrid,
    gridwidth=0.2pt,
    subgridwidth=0.1pt}

\psset{%
    linewidth=0.4pt
    }

\newcommand{\mypsgrid}{}
\newcommand{\mypslabel}{}
\newcommand{\mypsextra}{}

\newcommand{\mypsgenericnodes}{\psset{%
    radius=1.0mm,
    labelsep=2mm}}

\newcommand{\mypsgenericedges}{}
\newcommand{\mypsgenericedgesdashed}{\psset{linecolor=black,linestyle=dashed,dash=2pt 1pt}}

\newcommand{\mypsK}{\psset{%
    radius=1mm,
    labelsep=2mm}}

\newcommand{\mypsI}{\psset{%
    framesize=2mm,
    labelsep=2mm}}

\newcommand{\mypsV}{\psset{%
    radius=0.6mm,
    labelsep=1.5mm,
    fillstyle=solid,
    fillcolor=black}}

\newcommand{\mypsVinvisible}{}

\newcommand{\mypsv}[7]{%
    \Cnode(#1,#2){#3}%
    \ncarc[arcangle=#6]{-}{#3}{#4}%
    \ncarc[arcangle=#7]{-}{#3}{#5}%
}

\newcommand{\mypsvinvisible}[5]{%
    \pnode(#1,#2){#3}%
    \psset{fillstyle=none,linestyle=dashed,dash=2pt 1pt}
    \ncarc[arcangle=#5]{-}{#3}{#4}%
}

\newcommand{\mypsk}[2]{%
    \Cnode(#1,#2)%
}

\newcommand{\mypsi}[2]{%
    \fnode(#1,#2)%
}

\newcommand{\mynatural}{\mathbb{N}}
\newcommand{\myposinteger}{\mathbb{Z}^{+}}
\newcommand{\mysize}[1]{{\lvert #1 \rvert}}

\newcommand{\myceil}[1]{\lceil #1 \rceil}

\newcommand{\myA}{\mathcal{A}}
\newcommand{\myG}{\mathcal{G}}
\newcommand{\myH}{\mathcal{H}}
\newcommand{\myQ}{\mathcal{Q}}
\newcommand{\myT}{\mathcal{T}}

\newcommand{\myTvr}{\myT_v(r)}

\newcommand{\myIB}{I'}
\newcommand{\myKB}{K'}
\newcommand{\myEB}{E'}

\newcommand{\myt}[1]{\bar{#1}}
\newcommand{\myti}{\myt \iota}
\newcommand{\mytk}{\myt \kappa}

\newcommand{\myopt}[1]{#1^{*}}
\newcommand{\myutil}{\omega}
\newcommand{\myoptutil}{\myopt \myutil}

\newcommand{\Dk}{\Delta_K}
\newcommand{\Di}{\Delta_I}

\newcommand{\myPK}{D_K}
\newcommand{\myPI}{D_I}
\newcommand{\mylength}{s}
\newcommand{\mygirth}{g}

\newcommand{\myeqoverset}[2]{\overset{\text{\raisebox{1pt}{\makebox[0pt][c]{\eqref{#1}}}}}{#2}}

\newcommand{\myinst}{{\mathcal{S}}}

\newcommand{\myGkL}{\myG(k,L)}
\newcommand{\mykL}{kL}
\newcommand{\myhL}{hL}
\newcommand{\myaspacing}{a^{\vphantom{\mykL}}}
\newcommand{\mycspacing}{c^{\vphantom{\mykL}}}

\begin{document}
\title{Tight local approximation results \\ for max-min linear programs}
\author{Patrik Floréen \and Marja Hassinen \and Petteri Kaski \and Jukka Suomela}
\institute{%
    Helsinki Institute for Information Technology HIIT \\
    Department of Computer Science, University of Helsinki \\
    P.O. Box 68, FI-00014 University of Helsinki, Finland \\
    \email{$\{$firstname.lastname$\}$@cs.helsinki.fi}
}
\maketitle

\begin{abstract}
    In a bipartite max-min LP, we are given a bipartite graph $\myG = (V \cup I \cup K, E)$, where each agent $v \in V$ is adjacent to exactly one constraint $i \in I$ and exactly one objective $k \in K$. Each agent $v$ controls a variable $x_v$. For each $i \in I$ we have a nonnegative linear constraint on the variables of adjacent agents. For each $k \in K$ we have a nonnegative linear objective function of the variables of adjacent agents. The task is to maximise the minimum of the objective functions. We study local algorithms where each agent $v$ must choose $x_v$ based on input within its constant-radius neighbourhood in $\myG$. We show that for every $\epsilon>0$ there exists a local algorithm achieving the approximation ratio ${\Delta_I (1 - 1/\Delta_K)} + \epsilon$. We also show that this result is the best possible -- no local algorithm can achieve the approximation ratio ${\Delta_I (1 - 1/\Delta_K)}$. Here $\Delta_I$ is the maximum degree of a vertex $i \in I$, and $\Delta_K$ is the maximum degree of a vertex $k \in K$. As a methodological contribution, we introduce the technique of graph unfolding for the design of local approximation algorithms.
\end{abstract}

\section{Introduction}

As a motivating example, consider the task of data gathering in the following sensor network.
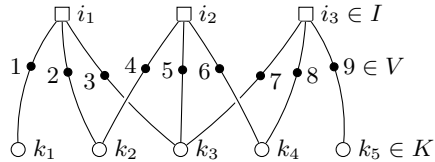
\begin{center}
    \newcommand{\mypslinks}{\psset{%
        border=1pt}}

    \newcommand{\mync}[4]{%
        {\mypslinks\ncarc[arcangle=#3]{-}{#1}{#2}}%
        {\mypsV\ncput{\Cnode(0,0){#4}}}%
    }

    \begin{pspicture}(0.4,0.8)(6.6,3.2)%
        \small%
        \mypsgrid
        {%
            \mypsK%
            \Cnode(0.6,1){k1}%
            \Cnode(1.8,1){k2}%
            \Cnode(3.0,1){k3}%
            \Cnode(4.2,1){k4}%
            \Cnode(5.4,1){k5}%
            \uput[r](k1){$k_1$}%
            \uput[r](k2){$k_2$}%
            \uput[r](k3){$k_3$}%
            \uput[r](k4){$k_4$}%
            \uput[r](k5){$k_5 \in K$}%
        }{%
            \mypsI%
            \fnode(1.25,3){i1}%
            \fnode(3.05,3){i2}%
            \fnode(4.85,3){i3}%
            \uput[r](i1){$i_1$}%
            \uput[r](i2){$i_2$}%
            \uput[r](i3){$i_3 \in I$}%
        }{%
            \psset{npos=0.6}%
            \mync{k1}{i1}{20}{v1}%
            \mync{k2}{i1}{15}{v2}%
            \mync{k3}{i1}{10}{v3}%
            \mync{k3}{i3}{-10}{v7}%
            \mync{k4}{i3}{-15}{v8}%
            \mync{k5}{i3}{-20}{v9}%
            \mync{k2}{i2}{5}{v4}%
            \mync{k3}{i2}{0}{v5}%
            \mync{k4}{i2}{-5}{v6}%
            \mypsV%
            \psset{labelsep=1.2mm}%
            \uput[180](v1){$1$}%
            \uput[190](v2){$2$}%
            \uput[200](v3){$3$}%
            \uput[170](v4){$4$}%
            \uput[180](v5){$5$}%
            \uput[190](v6){$6$}%
            \uput[-20](v7){$7$}%
            \uput[-10](v8){$8$}%
            \uput[0](v9){$9 \in V$}%
        }%
    \end{pspicture}
\end{center}
Each open circle is a sensor node $k \in K$, and each box is a relay node $i \in I$. The graph depicts the communication links between sensors and relays. Each sensor produces data which needs to be routed via adjacent relay nodes to a base station (not shown in the figure).

For each pair consisting of a sensor $k$ and an adjacent relay $i$, we need to decide how much data is routed from $k$ via $i$ to the base station. For each such decision, we introduce an \emph{agent} $v \in V$; these are shown as black dots in the figure. We arrive at a bipartite graph $\myG$ where the set of vertices is $V \cup I \cup K$ and each edge joins an agent $v \in V$ to a node $j \in I \cup K$.

Associated with each agent $v \in V$ is a variable $x_v$. Each relay constitutes a bottleneck: the relay has a limited battery capacity, which sets a limit on the total amount of data that can be forwarded through it. The task is to maximise the minimum amount of data gathered from a sensor node. In our example, the variable $x_2$ is the amount of data routed from the sensor $k_2$ via the relay $i_1$, the battery capacity of the relay $i_1$ is an upper bound for $x_1 + x_2 + x_3$, and the amount of data gathered from the sensor node $k_2$ is $x_2 + x_4$. Assuming that the maximum capacity of a relay is $1$, the optimisation problem is to
\begin{equation}
    \begin{alignedat}{2}
        &\text{maximise } \ &&{\min \ \{ x_1, \ x_2+x_4, \ x_3+x_5+x_7, \ x_6+x_8, \ x_9 \}} \\
        &\text{subject to } \ && x_1 + x_2 + x_3 \le 1, \\
        &&& x_4 + x_5 + x_6 \le 1, \\
        &&& x_7 + x_8 + x_9 \le 1, \\
        &&& x_1, x_2, \dotsc, x_9 \ge 0.
    \end{alignedat}\label{eq:sensor}
\end{equation}

In this work, we study \emph{local algorithms}~\cite{naor95what} for solving max-min linear programs (LPs) such as~\eqref{eq:sensor}. In a local algorithm, each agent $v \in V$ must choose the value $x_v$ solely based on its constant-radius neighbourhood in the graph~$\myG$. Such algorithms provide an extreme form of scalability in distributed systems; among others, a change in the topology of $\myG$ affects the values $x_v$ only in a constant-radius neighbourhood.

\subsection{Max-min linear programs}\label{ssec:max-min}

Let $\myG = (V \cup I \cup K, E)$ be a bipartite, undirected communication graph where each edge $e \in E$ is of the form $\{v, j\}$ with $v \in V$ and $j \in I \cup K$. The elements $v \in V$ are called \emph{agents}, the elements $i \in I$ are called \emph{constraints}, and the elements $k \in K$ are called \emph{objectives}; the sets $V$, $I$, and $K$ are disjoint. We define $V_i = {\{ v \in V : \{v,i\} \in E \}}$, $V_k = {\{ v \in V : \{v,k\} \in E \}}$, $I_v = {\{ i \in I : \{v,i\} \in E\}}$, and $K_v = {\{ k \in K : \{v,k\} \in E\}}$ for all $i \in I$, $k \in K$, $v \in V$.

We assume that $\myG$ is a bounded-degree graph; in particular, we assume that $\mysize{V_i} \le \Di$ and $\mysize{V_k} \le \Dk$ for all $i \in I$ and $k \in K$
for some constants $\Di$ and~$\Dk$.

A {\em max-min linear program} associated with $\myG$ is defined as follows. Associate a variable $x_v$ with each agent $v \in V$, associate a coefficient $a_{iv} \ge 0$ with each edge $\{i,v\} \in E$, $i \in I$, $v \in V$, and associate a coefficient $c_{kv} \ge 0$ with each edge $\{k,v\} \in E$, $k \in K$, $v \in V$. The task is to
\begin{equation}
    \begin{aligned}
        &\text{maximise } & \myutil \,=\, \textstyle\min_{k \in K} \textstyle\sum_{v \in V_k} c_{kv} x_v & \\
        &\text{subject to } & \textstyle\sum_{v \in V_i} a_{iv} x_v \,&\le\, 1 & \quad\forall\, &i \in I, \\
        && x_v \,&\ge\, 0 & \forall\, &v \in V .
    \end{aligned}\label{eq:max-min}
\end{equation}
We write $\myoptutil$ for the optimum of~\eqref{eq:max-min}.

\subsection{Special cases of max-min LPs}

A max-min LP is a generalisation of a \emph{packing LP}. Namely, in a packing LP there is only one linear nonnegative function to maximise, while in a max-min LP the goal is to maximise the minimum of multiple nonnegative linear functions.

Our main focus is on the \emph{bipartite version} of the max-min LP problem. In the bipartite version we have $|I_v| = |K_v| = 1$ for each $v \in V$. We also define the \emph{$0/1$ version}~\cite{floreen08local}. In that case we have $a_{iv} = 1$ and $c_{kv} = 1$ for all $v \in V, i \in I_v, k \in K_v$. Our example~\eqref{eq:sensor} is both a bipartite max-min LP and a 0/1 max-min LP.

The \emph{distance} between a pair of vertices $s, t \in V \cup I \cup K$ in $\myG$ is the number of edges on a shortest path connecting $s$ and $t$ in $\myG$. We write $B_\myG(s, r)$ for the set of vertices within distance at most $r$ from $s$. We say that $\myG$ has \emph{bounded relative growth} $1 + \delta$ \emph{beyond radius} $R \in \mynatural$ if ${\mysize{V \cap B_\myG(v, r+2)}} / {\mysize{V \cap B_\myG(v, r)}} \le {1 + \delta}$ for all $v \in V$, $r \ge R$. Any bounded-degree graph $\myG$ has a constant upper bound for $\delta$. Regular grids are a simple example of a family of graphs where $\delta$ approaches $0$ as $R$ increases~\cite{floreen08approximating}.

\subsection{Local algorithms and the model of computation}\label{ssec:model}

A local algorithm~\cite{naor95what} is a distributed algorithm in which the output of a node is a function of input available within a fixed-radius neighbourhood; put otherwise, the algorithm runs in a constant number of communication rounds. In the context of distributed max-min LPs, the exact definition is as follows.

We say that the \emph{local input} of a node $v \in V$ consists of the sets $I_v$ and $K_v$ and the coefficients $a_{iv}, c_{kv}$ for all $i \in I_v, k \in K_v$. The local input of a node $i \in I$ consists of $V_i$ and the local input of a node $k \in K$ consists of $V_k$. Furthermore, we assume that either (a) each node has a \emph{unique identifier} given as part of the local input to the node~\cite{naor95what,linial92locality}; or, (b) each vertex independently introduces an ordering of the edges incident to it. The latter, strictly weaker, assumption is often called \emph{port numbering}~\cite{angluin80local}; in essence, each edge $\{s,t\}$ in $\myG$ has two natural numbers associated with it: the port number in $s$ and the port number in $t$.

Let $\myA$ be a deterministic distributed algorithm executed by each of the nodes of $\myG$ that finds a feasible solution $x$ to any max-min LP~\eqref{eq:max-min} given locally as input to the nodes. Let $r \in \mynatural$ be a constant independent of the input. We say that $\myA$ is a \emph{local algorithm} with \emph{local horizon} $r$ if, for every agent $v \in V$, the output $x_v$ is a function of the local input of the nodes in $B_\myG(v, r)$. Furthermore, we say that $\myA$ has the \emph{approximation ratio} $\alpha \ge 1$ if $\sum_{v \in V_k} c_{kv} x_v \ge \myoptutil / \alpha$ for all $k \in K$.

\subsection{Contributions and prior work}

The following local approximability result is the main contribution of this paper.

\begin{theorem}\label{thm:approx}
    For any $\Di \ge 2$, $\Dk \ge 2$, and $\epsilon > 0$, there exists a local approximation algorithm for the bipartite max-min LP problem with the approximation ratio ${\Di (1 - 1/\Dk) + \epsilon}$. The algorithm assumes only port numbering.
\end{theorem}

We also show that the positive result of Theorem~\ref{thm:approx} is tight. Namely, we prove a matching lower bound on local approximability, which holds even if we assume both 0/1 coefficients and unique node identifiers given as input.

\begin{theorem}\label{thm:inapprox}
    For any $\Di \ge 2$ and $\Dk \ge 2$, there exists no local approximation algorithm for the max-min LP problem with the approximation ratio ${\Di (1 - 1/\Dk)}$. This holds even in the case of a bipartite, $0/1$ max-min LP and with unique node identifiers given as input.
\end{theorem}

Considering Theorem~\ref{thm:approx} in light of Theorem~\ref{thm:inapprox}, we find it somewhat surprising that unique node identifiers are not required to obtain the best possible local approximation algorithm for bipartite max-min LPs.

In terms of earlier work, Theorem~\ref{thm:approx} is an improvement on the \emph{safe algorithm}~\cite{floreen08approximating,papadimitriou93linear} which achieves the approximation ratio $\Di$. Theorem~\ref{thm:inapprox} improves upon the earlier lower bound ${(\Di+1)/2} - {1/(2\Dk-2)}$~\cite{floreen08approximating}; here it should be noted that our definition of the local horizon differs by a constant factor from earlier work~\cite{floreen08approximating} due to the fact that we have adopted a more convenient graph representation instead of a hypergraph representation. 

In the context of packing and covering LPs, it is known~\cite{kuhn06price} that any approximation ratio $\alpha > 1$ can be achieved by a local algorithm, assuming a bounded-degree graph and bounded coefficients. Compared with this, the factor ${\Di (1 - 1/\Dk)}$ approximation in Theorem~\ref{thm:approx} sounds somewhat discouraging considering practical applications. However, the constructions that we use in our negative results are arguably far from the structure of, say, a typical real-world wireless network. In prior work~\cite{floreen08approximating} we presented a local algorithm that achieves a factor $1 + (2+o(1))\delta$ approximation assuming that $\myG$ has bounded relative growth $1+\delta$ beyond some constant radius $R$; for a small $\delta$, this is considerably better than ${\Di (1 - 1/\Dk)}$ for general graphs. We complement this line of research on bounded relative growth graphs with a negative result that matches the prior positive result~\cite{floreen08approximating} up to constants.

\begin{theorem}\label{thm:inapprox-bounded}
    Let $\Di \ge 3$, $\Dk \ge 3$, and $0 < \delta < 1/10$. There exists no local approximation algorithm for the max-min LP problem with an approximation ratio less than ${1 + \delta/2}$. This holds even in the case of a bipartite max-min LP where the graph $\myG$ has bounded relative growth $1+\delta$ beyond some constant radius~$R$.
\end{theorem}

From a technical perspective, the proof of Theorem~\ref{thm:approx} relies on two ideas: \emph{graph unfolding} and the idea of \emph{averaging local solutions} of local LPs.

We introduce the unfolding technique in Sect.~\ref{sec:unfolding}. In essence, we expand the finite input graph~$\myG$ into a possibly infinite tree~$\myT$. Technically, $\myT$ is the \emph{universal covering} of $\myG$~\cite{angluin80local}. While such unfolding arguments have been traditionally used to obtain impossibility results~\cite{lynch89hundred} in the context of distributed algorithms, here we use such an argument to simplify the design of local algorithms. In retrospect, our earlier approximation algorithm for 0/1 max-min LPs~\cite{floreen08local} can be interpreted as an application of the unfolding technique.

The idea of averaging local LPs has been used commonly in prior work on distributed algorithms~\cite{floreen08approximating,kuhn06price,kuhn07distributed,kuhn05locality}. Our algorithm can also be interpreted as a generalisation of the safe algorithm~\cite{papadimitriou93linear} beyond local horizon $r = 1$.

To obtain our negative results -- Theorems \ref{thm:inapprox} and \ref{thm:inapprox-bounded} -- we use a construction based on regular high-girth graphs. Such graphs~\cite{lubotzky88ramanujan,lazebnik95explicit,hoory02graphs,mckay04short} have been used in prior work to obtain impossibility results related to local algorithms~\cite{linial92locality,kuhn06price,kuhn04what}.

\section{Graph unfolding}\label{sec:unfolding}

Let $\myH = (V, E)$ be a connected undirected graph and let $v \in V$. Construct a (possibly infinite) rooted tree $\myT_v = (\myt V, \myt E)$ and a labelling $f_v \colon \myt V \to V$ as follows. First, introduce a vertex $\myt v$ as the root of $\myT_v$ and set $f_v(\myt v) = v$. Then, for each vertex $u$ adjacent to $v$ in $\myH$, add a new vertex $\myt u$ as a child of $\myt v$ and set $f_v(\myt u) = u$. Then expand recursively as follows. For each unexpanded $\myt t\neq\myt v$ with parent $\myt s$, and each $u \ne f(\myt s)$ adjacent 
to $f(\myt t)$ in $\myH$, add a new vertex $\myt u$ as a child of 
$\myt t$ and set $f_v(\myt u) = u$. Mark $\myt t$ as expanded.

This construction is illustrated in Fig.~\ref{fig:unfolding}. Put simply, we traverse $\myH$ in a breadth-first manner and treat vertices revisited due to a cycle as new vertices; in particular, the tree $\myT_v$ is finite if and only if $\myH$ is acyclic.
\begin{figure}[b]
    \centering
    \begin{pspicture}(0.0,-0.5)(2.2,3.0)%
        \mypsgrid
        \rput[lB](0.1,2.7){$\myH$:}
        \mypsgenericnodes
        \Cnode(0.2, 0.6){b}
        \Cnode(0.2, 1.8){a}
        \Cnode(1.0, 1.2){c}
        \Cnode(1.8, 1.2){d}
        \uput[u](a){$a$}
        \uput[d](b){$b$}
        \uput[u](c){$c$}
        \uput[r](d){$d$}
        \mypsgenericedges
        \ncarc[arcangle=-5]{-}{c}{a}
        \ncarc[arcangle=5]{-}{c}{b}
        \ncarc[arcangle=10]{-}{c}{d}
        \ncarc[arcangle=-5]{-}{a}{b}
    \end{pspicture}
    \hspace{\stretch{1}}
    \begin{pspicture}(-0.8,-0.5)(2.2,3.0)%
        \mypsgrid
        \rput[lB](-0.8,2.7){$(\myT_a, f_a)$:}
        \mypsgenericnodes
        \Cnode(0.8, 2.0){a0}\uput[u](a0){$a$}
        \Cnode(0.0, 1.5){b1}\uput[l](b1){$b$}
        \Cnode(1.6, 1.5){c1}\uput[r](c1){$c$}
        \Cnode(0.0, 1.0){c2}\uput[l](c2){$c$}
        \Cnode(1.3, 1.0){b2}\uput[l](b2){$b$}
        \Cnode(1.9, 1.0){d2}\uput[d](d2){$d$}
        \Cnode(-0.3, 0.5){a3v1}\uput[l](a3v1){$a$}
        \Cnode(0.3, 0.5){d3v1}\uput[d](d3v1){$d$}
        \Cnode(1.3, 0.5){a3v2}\uput[l](a3v2){$a$}
        \Cnode(-0.3, 0.0){b4}\uput[l](b4){$b$}
        \Cnode(1.3, 0.0){c4}\uput[l](c4){$c$}
        \pnode(-0.3, -0.5){c5}
        \pnode(1.0, -0.5){b5}
        \pnode(1.6, -0.5){d5}
        \mypsgenericedges
        \ncarc[arcangle=-10]{-}{a0}{b1}
        \ncarc[arcangle=10]{-}{a0}{c1} 
        \ncarc[arcangle=-10]{-}{b1}{c2}
        \ncarc[arcangle=-10]{-}{c1}{b2}
        \ncarc[arcangle=10]{-}{c1}{d2}
        \ncarc[arcangle=-10]{-}{c2}{a3v1}
        \ncarc[arcangle=10]{-}{c2}{d3v1}
        \ncarc[arcangle=-10]{-}{b2}{a3v2} 
        \ncarc[arcangle=-10]{-}{a3v1}{b4}
        \ncarc[arcangle=-10]{-}{a3v2}{c4}
        \mypsgenericedgesdashed
        \ncarc[arcangle=-2]{-}{b4}{c5}
        \ncarc[arcangle=-10]{-}{c4}{b5}
        \ncarc[arcangle=10]{-}{c4}{d5}
    \end{pspicture}
    \hspace{\stretch{1}}
    \begin{pspicture}(-0.8,-0.5)(2.3,3.0)%
        \mypsgrid
        \rput[lB](-0.6,2.7){$(\myT_c, f_c)$:}
        \mypsgenericnodes
        \Cnode(1.0, 2.0){c0}\uput[u](c0){$c$}
        \Cnode(0.0, 1.5){a1}\uput[l](a1){$a$}
        \Cnode(1.3, 1.5){b1}\uput[l](b1){$b$}
        \Cnode(2.0, 1.5){d1}\uput[r](d1){$d$}
        \Cnode(0.0, 1.0){b2}\uput[l](b2){$b$}
        \Cnode(1.3, 1.0){a2}\uput[l](a2){$a$}
        \Cnode(0.0, 0.5){c3v1}\uput[l](c3v1){$c$}
        \Cnode(1.3, 0.5){c3v2}\uput[l](c3v2){$c$}
        \Cnode(-0.3, 0.0){a4v1}\uput[l](a4v1){$a$}
        \Cnode(0.3, 0.0){d4v1}\uput[d](d4v1){$d$}
        \Cnode(1.0, 0.0){b4v2}\uput[l](b4v2){$b$}
        \Cnode(1.6, 0.0){d4v2}\uput[d](d4v2){$d$}
        \pnode(-0.3, -0.5){i0}
        \pnode(1.0, -0.5){i1}
        \mypsgenericedges
        \ncarc[arcangle=-10]{-}{c0}{a1}
        \ncarc[arcangle=10]{-}{c0}{b1}
        \ncarc[arcangle=10]{-}{c0}{d1}
        \ncarc[arcangle=-10]{-}{a1}{b2}
        \ncarc[arcangle=-10]{-}{b1}{a2}  
        \ncarc[arcangle=-10]{-}{b2}{c3v1}
        \ncarc[arcangle=-10]{-}{a2}{c3v2}
        \ncarc[arcangle=-10]{-}{c3v1}{a4v1}
        \ncarc[arcangle=10]{-}{c3v1}{d4v1}
        \ncarc[arcangle=-10]{-}{c3v2}{b4v2}
        \ncarc[arcangle=10]{-}{c3v2}{d4v2}
        \mypsgenericedgesdashed
        \ncarc[arcangle=-2]{-}{a4v1}{i0}
        \ncarc[arcangle=-2]{-}{b4v2}{i1}
    \end{pspicture}
    \hspace{\stretch{1}}
    \begin{pspicture}(-0.6,-0.5)(1.4,3.0)%
        \mypsgrid
        \rput[lB](-0.6,2.7){$(\myT,f)$:}
        \mypsgenericnodes
        \Cnode(1.0, 0.0){a0}\uput[r](a0){$a$}
        \Cnode(1.0, 0.5){c0}\uput[r](c0){$c$}
        \Cnode(1.0, 1.0){b0}\uput[r](b0){$b$}
        \Cnode(1.0, 1.5){a1}\uput[r](a1){$a$}
        \Cnode(1.0, 2.0){c1}\uput[r](c1){$c$}
        \Cnode(1.0, 2.5){b1}\uput[r](b1){$b$}
        \Cnode(0.5, 0.5){d0}\uput[l](d0){$d$}
        \Cnode(0.5, 2.0){d1}\uput[l](d1){$d$}
        \pnode(1.0, -0.5){i0}
        \pnode(1.0, 3.0){i1}
        \mypsgenericedges
        \ncarc[arcangle=-10]{-}{a0}{c0}
        \ncarc[arcangle=-10]{-}{c0}{b0}
        \ncarc[arcangle=-10]{-}{b0}{a1}
        \ncarc[arcangle=-10]{-}{a1}{c1}
        \ncarc[arcangle=-10]{-}{c1}{b1}
        \ncarc[arcangle=-10]{-}{c0}{d0}
        \ncarc[arcangle=-10]{-}{c1}{d1}
        \mypsgenericedgesdashed
        \ncarc[arcangle=-2]{-}{a0}{i0}
        \ncarc[arcangle=-2]{-}{b1}{i1}
    \end{pspicture}
    \caption{An example graph $\myH$ and its unfolding $(\myT,f)$.}\label{fig:unfolding}
\end{figure}

The rooted, labelled trees $(\myT_v,f_v)$ obtained in this way for different choices of $v\in V$ are isomorphic viewed as unrooted trees~\cite{angluin80local}. For example, the infinite labelled trees $(\myT_a, f_a)$ and $(\myT_c, f_c)$ in Fig.~\ref{fig:unfolding} are isomorphic and can be transformed into each other by rotations. Thus, we can define the \emph{unfolding} of $\myH$ as the labelled tree $(\myT, f)$ where $\myT$ is the unrooted version of $\myT_v$ and $f = f_v$; up to isomorphism, this is independent of the choice of $v \in V$. Appendix~\ref{app:terminology} provides a further discussion on the terminology and concepts related to unfolding. 

\subsection{Unfolding and local algorithms}\label{sec:unfold-local}

Let us now view the graph $\myH$ as the communication graph of a distributed system, and let $(\myT,f)$ be the unfolding of $\myH$. Even if $\myT$ in general is countably infinite, a local algorithm $\myA$ with local horizon $r$ can be designed to operate at a node of $v \in \myH$ exactly \emph{as if} it was a node $\myt v \in f^{-1}(v)$ in the communication graph~$\myT$. Indeed, assume that the local input at $\myt v$ is identical to the local input at $f(\myt v)$, and observe that the radius $r$ neighbourhood of the node $\myt v$ in $\myT$ is equal to the rooted tree $\myT_v$ trimmed to depth $r$; let us denote this by $\myTvr$. To gather the information in $\myTvr$, it is sufficient to gather information on all walks of length at most $r$ starting at $v$ in $\myH$; using port numbering, the agents can detect and discard walks that consecutively traverse the same edge.

Assuming that only port numbering is available, the information in $\myTvr$ is in fact \emph{all} that the agent $v$ can gather. Indeed, to assemble, say, the subgraph of $\myH$ induced by $B_\myH(v,r)$, the agent $v$ in general needs to distinguish between a short cycle and a long path, and these are indistinguishable without node identifiers.

\subsection{Unfolding and max-min LPs}\label{sec:unfold-maxmin}

Let us now consider a max-min LP associated with a graph $\myG$. The unfolding of $\myG$ leads in a natural way to the unfolding of the max-min LP. As the unfolding of a max-min LP is, in general, countably infinite, we need minor technical extensions reviewed in Appendix~\ref{app:infinite}. A formal definition of the unfolding of a max-min LP and the proof of the following lemma is given in Appendix \ref{app:unfolding}.

\begin{lemma}\label{lem:unfolding}
    Let $\myt\myA$ be a local algorithm for unfoldings of a family of max-min LPs and let $\alpha \ge 1$. Assume that the output $x$ of $\myt\myA$ satisfies $\sum_{v \in V_k} c_{kv} x_v \ge \myutil' / \alpha$ for all $k \in K$ if there exists a feasible solution with utility at least $\myutil'$. Furthermore, assume that $\myt\myA$ uses port numbering only. Then, there exists a local approximation algorithm $\myA$ with the approximation ratio $\alpha$ for this family of max-min LPs.
\end{lemma}

\section{Approximability results}\label{sec:approx}

We proceed to prove Theorem~\ref{thm:approx}. 
Let $\Di\geq 2$, $\Dk\geq 2$, and $\epsilon>0$ be fixed.
By virtue of Lemma~\ref{lem:unfolding}, it suffices to consider only
bipartite max-min LPs where the graph $\myG$ is a (finite or countably infinite) tree.

To ease the analysis, it will be convenient to \emph{regularise} $\myG$ to a countably infinite tree with $\mysize{V_i} = \Di$ and $\mysize{V_k} = \Dk$ for all $i \in I$ and $k \in K$.

To this end, if $\mysize{V_i} < \Di$ for some $i \in I$, add $\Di - \mysize{V_i}$ new \emph{virtual} agents as neighbours of $i$. Let $v$ be one of these agents. Set $a_{iv} = 0$ so that no matter what value one assigns to $x_v$, it does not affect the feasibility of the constraint $i$. Then add a new virtual objective $k$ adjacent to $v$ and set, for example, $c_{kv} = 1$. As one can assign an arbitrarily large value to $x_v$, the virtual objective $k$ will not be a bottleneck.

Similarly, if $\mysize{V_k} < \Dk$ for some $k \in K$, add $\Dk - \mysize{V_k}$ new virtual agents as neighbours of $k$. Let $v$ be one of these agents. Set $c_{kv} = 0$ so that no matter what value one assigns to $x_v$, it does not affect the value of the objective $k$. Then add a new virtual constraint $i$ adjacent to $v$ and set, for example, $a_{iv} = 1$.  

\begin{figure}[t]
    \centering
    \renewcommand{\mypslabel}{(a)}
    \renewcommand{\mypsextra}{%
        \uput[0](k0){$k_0$}%
        \uput[-90](a2){$v_0$}%
    }
    \begin{pspicture}(-2.17,-2.17)(2.17,2.17)%
    \mypsgrid
    \rput[lB](-1.97,-1.97){\mypslabel}
    \mypsK
    \mypsk{0}{0}{k0}
    \mypsk{-0.18855}{-1.2104}{k1}
    \mypsk{0.633594}{-1.04842}{k2}
    \mypsk{1.15927}{-0.39587}{k3}
    \mypsk{1.14251}{0.441912}{k4}
    \mypsk{0.591161}{1.07292}{k5}
    \mypsk{-0.236803}{1.20189}{k6}
    \mypsk{-0.953964}{0.768491}{k7}
    \mypsk{-1.22476}{-0.0244984}{k8}
    \mypsk{-0.92247}{-0.806024}{k9}
    \mypsI
    \mypsi{0.362054}{-0.599097}{i0}
    \mypsi{0.337806}{0.613096}{i1}
    \mypsi{-0.69986}{-0.0139991}{i2}
    \mypsi{-0.565528}{-1.6561}{i3}
    \mypsi{0.0349977}{-1.74965}{i4}
    \mypsi{0.631303}{-1.63216}{i5}
    \mypsi{1.15146}{-1.31781}{i6}
    \mypsi{1.53274}{-0.844516}{i7}
    \mypsi{1.72915}{-0.269358}{i8}
    \mypsi{1.71699}{0.33829}{i9}
    \mypsi{1.49774}{0.905134}{i10}
    \mypsi{1.09784}{1.36281}{i11}
    \mypsi{0.565528}{1.6561}{i12}
    \mypsi{-0.0349977}{1.74965}{i13}
    \mypsi{-0.631303}{1.63216}{i14}
    \mypsi{-1.15146}{1.31781}{i15}
    \mypsi{-1.53274}{0.844516}{i16}
    \mypsi{-1.72915}{0.269358}{i17}
    \mypsi{-1.71699}{-0.33829}{i18}
    \mypsi{-1.49774}{-0.905134}{i19}
    \mypsi{-1.09784}{-1.36281}{i20}
    \mypsV
    \mypsv{0.181027}{-0.299549}{a0}{k0}{i0}{0}{0}
    \mypsv{0.168903}{0.306548}{a1}{k0}{i1}{0}{0}
    \mypsv{-0.34993}{-0.00699953}{a2}{k0}{i2}{0}{0}
    \mypsv{0.0867516}{-0.90475}{a3}{k1}{i0}{0}{0}
    \mypsv{0.497824}{-0.823758}{a4}{k2}{i0}{0}{0}
    \mypsv{0.760663}{-0.497483}{a5}{k3}{i0}{0}{0}
    \mypsv{0.74016}{0.527504}{a6}{k4}{i1}{0}{0}
    \mypsv{0.464484}{0.843007}{a7}{k5}{i1}{0}{0}
    \mypsv{0.0505019}{0.907495}{a8}{k6}{i1}{0}{0}
    \mypsv{-0.826912}{0.377246}{a9}{k7}{i2}{0}{0}
    \mypsv{-0.962308}{-0.0192487}{a10}{k8}{i2}{0}{0}
    \mypsv{-0.811165}{-0.410012}{a11}{k9}{i2}{0}{0}
    \mypsv{-0.377039}{-1.43325}{a12}{k1}{i3}{0}{0}
    \mypsv{-0.0767763}{-1.48003}{a13}{k1}{i4}{0}{0}
    \mypsv{0.632448}{-1.34029}{a14}{k2}{i5}{0}{0}
    \mypsv{0.892528}{-1.18312}{a15}{k2}{i6}{0}{0}
    \mypsv{1.34601}{-0.620193}{a16}{k3}{i7}{0}{0}
    \mypsv{1.44421}{-0.332614}{a17}{k3}{i8}{0}{0}
    \mypsv{1.42975}{0.390101}{a18}{k4}{i9}{0}{0}
    \mypsv{1.32013}{0.673523}{a19}{k4}{i10}{0}{0}
    \mypsv{0.844502}{1.21786}{a20}{k5}{i11}{0}{0}
    \mypsv{0.578345}{1.36451}{a21}{k5}{i12}{0}{0}
    \mypsv{-0.1359}{1.47577}{a22}{k6}{i13}{0}{0}
    \mypsv{-0.434053}{1.41703}{a23}{k6}{i14}{0}{0}
    \mypsv{-1.05271}{1.04315}{a24}{k7}{i15}{0}{0}
    \mypsv{-1.24335}{0.806503}{a25}{k7}{i16}{0}{0}
    \mypsv{-1.47695}{0.12243}{a26}{k8}{i17}{0}{0}
    \mypsv{-1.47087}{-0.181394}{a27}{k8}{i18}{0}{0}
    \mypsv{-1.21011}{-0.855579}{a28}{k9}{i19}{0}{0}
    \mypsv{-1.01016}{-1.08441}{a29}{k9}{i20}{0}{0}
    \mypsVinvisible
    \mypsvinvisible{-0.810541}{-1.93353}{a30}{i3}{0}
    \mypsvinvisible{-0.678634}{-1.98732}{a31}{i3}{0}
    \mypsvinvisible{-0.541374}{-2.02544}{a32}{i3}{0}
    \mypsvinvisible{-0.100354}{-2.09414}{a33}{i4}{0}
    \mypsvinvisible{0.0419972}{-2.09958}{a34}{i4}{0}
    \mypsvinvisible{0.184017}{-2.08845}{a35}{i4}{0}
    \mypsvinvisible{0.621937}{-2.00217}{a36}{i5}{0}
    \mypsvinvisible{0.757563}{-1.9586}{a37}{i5}{0}
    \mypsvinvisible{0.887213}{-1.89957}{a38}{i5}{0}
    \mypsvinvisible{1.26921}{-1.66871}{a39}{i6}{0}
    \mypsvinvisible{1.38176}{-1.58138}{a40}{i6}{0}
    \mypsvinvisible{1.4834}{-1.48157}{a41}{i6}{0}
    \mypsvinvisible{1.7634}{-1.13398}{a42}{i7}{0}
    \mypsvinvisible{1.83929}{-1.01342}{a43}{i7}{0}
    \mypsvinvisible{1.90066}{-0.884864}{a44}{i7}{0}
    \mypsvinvisible{2.0449}{-0.462473}{a45}{i8}{0}
    \mypsvinvisible{2.07498}{-0.323229}{a46}{i8}{0}
    \mypsvinvisible{2.08868}{-0.181435}{a47}{i8}{0}
    \mypsvinvisible{2.07975}{0.264815}{a48}{i9}{0}
    \mypsvinvisible{2.06039}{0.405947}{a49}{i9}{0}
    \mypsvinvisible{2.02477}{0.543878}{a50}{i9}{0}
    \mypsvinvisible{1.86376}{0.960162}{a51}{i10}{0}
    \mypsvinvisible{1.79729}{1.08616}{a52}{i10}{0}
    \mypsvinvisible{1.71665}{1.20359}{a53}{i10}{0}
    \mypsvinvisible{1.42296}{1.5397}{a54}{i11}{0}
    \mypsvinvisible{1.31741}{1.63537}{a55}{i11}{0}
    \mypsvinvisible{1.20147}{1.71813}{a56}{i11}{0}
    \mypsvinvisible{0.810541}{1.93353}{a57}{i12}{0}
    \mypsvinvisible{0.678634}{1.98732}{a58}{i12}{0}
    \mypsvinvisible{0.541374}{2.02544}{a59}{i12}{0}
    \mypsvinvisible{0.100354}{2.09414}{a60}{i13}{0}
    \mypsvinvisible{-0.0419972}{2.09958}{a61}{i13}{0}
    \mypsvinvisible{-0.184017}{2.08845}{a62}{i13}{0}
    \mypsvinvisible{-0.621937}{2.00217}{a63}{i14}{0}
    \mypsvinvisible{-0.757563}{1.9586}{a64}{i14}{0}
    \mypsvinvisible{-0.887213}{1.89957}{a65}{i14}{0}
    \mypsvinvisible{-1.26921}{1.66871}{a66}{i15}{0}
    \mypsvinvisible{-1.38176}{1.58138}{a67}{i15}{0}
    \mypsvinvisible{-1.4834}{1.48157}{a68}{i15}{0}
    \mypsvinvisible{-1.7634}{1.13398}{a69}{i16}{0}
    \mypsvinvisible{-1.83929}{1.01342}{a70}{i16}{0}
    \mypsvinvisible{-1.90066}{0.884864}{a71}{i16}{0}
    \mypsvinvisible{-2.0449}{0.462473}{a72}{i17}{0}
    \mypsvinvisible{-2.07498}{0.323229}{a73}{i17}{0}
    \mypsvinvisible{-2.08868}{0.181435}{a74}{i17}{0}
    \mypsvinvisible{-2.07975}{-0.264815}{a75}{i18}{0}
    \mypsvinvisible{-2.06039}{-0.405947}{a76}{i18}{0}
    \mypsvinvisible{-2.02477}{-0.543878}{a77}{i18}{0}
    \mypsvinvisible{-1.86376}{-0.960162}{a78}{i19}{0}
    \mypsvinvisible{-1.79729}{-1.08616}{a79}{i19}{0}
    \mypsvinvisible{-1.71665}{-1.20359}{a80}{i19}{0}
    \mypsvinvisible{-1.42296}{-1.5397}{a81}{i20}{0}
    \mypsvinvisible{-1.31741}{-1.63537}{a82}{i20}{0}
    \mypsvinvisible{-1.20147}{-1.71813}{a83}{i20}{0}
    \mypsextra
    \end{pspicture}
    \hspace{1cm}
    \renewcommand{\mypslabel}{(b)}
    \renewcommand{\mypsextra}{%
        \uput[-45](i0){$i_0$}%
    }
    \begin{pspicture}(-2.17,-2.17)(2.17,2.17)%
    \mypsgrid
    \rput[lB](-1.97,-1.97){\mypslabel}
    \mypsK
    \mypsk{0.0139991}{-0.69986}{k0}
    \mypsk{0.69986}{0.0139991}{k1}
    \mypsk{-0.0139991}{0.69986}{k2}
    \mypsk{-0.69986}{-0.0139991}{k3}
    \mypsk{-1.03735}{-1.4094}{k4}
    \mypsk{-0.637228}{-1.62986}{k5}
    \mypsk{-0.193677}{-1.73925}{k6}
    \mypsk{0.263073}{-1.73011}{k7}
    \mypsk{0.701896}{-1.60307}{k8}
    \mypsk{1.09288}{-1.36679}{k9}
    \mypsk{1.4094}{-1.03735}{k10}
    \mypsk{1.62986}{-0.637228}{k11}
    \mypsk{1.73925}{-0.193677}{k12}
    \mypsk{1.73011}{0.263073}{k13}
    \mypsk{1.60307}{0.701896}{k14}
    \mypsk{1.36679}{1.09288}{k15}
    \mypsk{1.03735}{1.4094}{k16}
    \mypsk{0.637228}{1.62986}{k17}
    \mypsk{0.193677}{1.73925}{k18}
    \mypsk{-0.263073}{1.73011}{k19}
    \mypsk{-0.701896}{1.60307}{k20}
    \mypsk{-1.09288}{1.36679}{k21}
    \mypsk{-1.4094}{1.03735}{k22}
    \mypsk{-1.62986}{0.637228}{k23}
    \mypsk{-1.73925}{0.193677}{k24}
    \mypsk{-1.73011}{-0.263073}{k25}
    \mypsk{-1.60307}{-0.701896}{k26}
    \mypsk{-1.36679}{-1.09288}{k27}
    \mypsI
    \mypsi{0}{0}{i0}
    \mypsi{-0.44606}{-1.1409}{i1}
    \mypsi{0.491327}{-1.12215}{i2}
    \mypsi{1.1409}{-0.44606}{i3}
    \mypsi{1.12215}{0.491327}{i4}
    \mypsi{0.44606}{1.1409}{i5}
    \mypsi{-0.491327}{1.12215}{i6}
    \mypsi{-1.1409}{0.44606}{i7}
    \mypsi{-1.12215}{-0.491327}{i8}
    \mypsV
    \mypsv{0.00699953}{-0.34993}{a0}{k0}{i0}{0}{0}
    \mypsv{0.34993}{0.00699953}{a1}{k1}{i0}{0}{0}
    \mypsv{-0.00699953}{0.34993}{a2}{k2}{i0}{0}{0}
    \mypsv{-0.34993}{-0.00699953}{a3}{k3}{i0}{0}{0}
    \mypsv{-0.21603}{-0.920381}{a4}{k0}{i1}{0}{0}
    \mypsv{0.252663}{-0.911005}{a5}{k0}{i2}{0}{0}
    \mypsv{0.920381}{-0.21603}{a6}{k1}{i3}{0}{0}
    \mypsv{0.911005}{0.252663}{a7}{k1}{i4}{0}{0}
    \mypsv{0.21603}{0.920381}{a8}{k2}{i5}{0}{0}
    \mypsv{-0.252663}{0.911005}{a9}{k2}{i6}{0}{0}
    \mypsv{-0.920381}{0.21603}{a10}{k3}{i7}{0}{0}
    \mypsv{-0.911005}{-0.252663}{a11}{k3}{i8}{0}{0}
    \mypsv{-0.741707}{-1.27515}{a12}{k4}{i1}{0}{0}
    \mypsv{-0.541644}{-1.38538}{a13}{k5}{i1}{0}{0}
    \mypsv{-0.319868}{-1.44008}{a14}{k6}{i1}{0}{0}
    \mypsv{0.3772}{-1.42613}{a15}{k7}{i2}{0}{0}
    \mypsv{0.596611}{-1.36261}{a16}{k8}{i2}{0}{0}
    \mypsv{0.792106}{-1.24447}{a17}{k9}{i2}{0}{0}
    \mypsv{1.27515}{-0.741707}{a18}{k10}{i3}{0}{0}
    \mypsv{1.38538}{-0.541644}{a19}{k11}{i3}{0}{0}
    \mypsv{1.44008}{-0.319868}{a20}{k12}{i3}{0}{0}
    \mypsv{1.42613}{0.3772}{a21}{k13}{i4}{0}{0}
    \mypsv{1.36261}{0.596611}{a22}{k14}{i4}{0}{0}
    \mypsv{1.24447}{0.792106}{a23}{k15}{i4}{0}{0}
    \mypsv{0.741707}{1.27515}{a24}{k16}{i5}{0}{0}
    \mypsv{0.541644}{1.38538}{a25}{k17}{i5}{0}{0}
    \mypsv{0.319868}{1.44008}{a26}{k18}{i5}{0}{0}
    \mypsv{-0.3772}{1.42613}{a27}{k19}{i6}{0}{0}
    \mypsv{-0.596611}{1.36261}{a28}{k20}{i6}{0}{0}
    \mypsv{-0.792106}{1.24447}{a29}{k21}{i6}{0}{0}
    \mypsv{-1.27515}{0.741707}{a30}{k22}{i7}{0}{0}
    \mypsv{-1.38538}{0.541644}{a31}{k23}{i7}{0}{0}
    \mypsv{-1.44008}{0.319868}{a32}{k24}{i7}{0}{0}
    \mypsv{-1.42613}{-0.3772}{a33}{k25}{i8}{0}{0}
    \mypsv{-1.36261}{-0.596611}{a34}{k26}{i8}{0}{0}
    \mypsv{-1.24447}{-0.792106}{a35}{k27}{i8}{0}{0}
    \mypsVinvisible
    \mypsvinvisible{-1.3078}{-1.64167}{a36}{k4}{0}
    \mypsvinvisible{-1.17874}{-1.73666}{a37}{k4}{0}
    \mypsvinvisible{-0.838338}{-1.92421}{a38}{k5}{0}
    \mypsvinvisible{-0.689101}{-1.98256}{a39}{k5}{0}
    \mypsvinvisible{-0.311749}{-2.07563}{a40}{k6}{0}
    \mypsvinvisible{-0.152495}{-2.09336}{a41}{k6}{0}
    \mypsvinvisible{0.236085}{-2.08559}{a42}{k7}{0}
    \mypsvinvisible{0.394502}{-2.0615}{a43}{k7}{0}
    \mypsvinvisible{0.767831}{-1.95342}{a44}{k8}{0}
    \mypsvinvisible{0.914615}{-1.88915}{a45}{k8}{0}
    \mypsvinvisible{1.24725}{-1.68813}{a46}{k9}{0}
    \mypsvinvisible{1.3724}{-1.58806}{a47}{k9}{0}
    \mypsvinvisible{1.64167}{-1.3078}{a48}{k10}{0}
    \mypsvinvisible{1.73666}{-1.17874}{a49}{k10}{0}
    \mypsvinvisible{1.92421}{-0.838338}{a50}{k11}{0}
    \mypsvinvisible{1.98256}{-0.689101}{a51}{k11}{0}
    \mypsvinvisible{2.07563}{-0.311749}{a52}{k12}{0}
    \mypsvinvisible{2.09336}{-0.152495}{a53}{k12}{0}
    \mypsvinvisible{2.08559}{0.236085}{a54}{k13}{0}
    \mypsvinvisible{2.0615}{0.394502}{a55}{k13}{0}
    \mypsvinvisible{1.95342}{0.767831}{a56}{k14}{0}
    \mypsvinvisible{1.88915}{0.914615}{a57}{k14}{0}
    \mypsvinvisible{1.68813}{1.24725}{a58}{k15}{0}
    \mypsvinvisible{1.58806}{1.3724}{a59}{k15}{0}
    \mypsvinvisible{1.3078}{1.64167}{a60}{k16}{0}
    \mypsvinvisible{1.17874}{1.73666}{a61}{k16}{0}
    \mypsvinvisible{0.838338}{1.92421}{a62}{k17}{0}
    \mypsvinvisible{0.689101}{1.98256}{a63}{k17}{0}
    \mypsvinvisible{0.311749}{2.07563}{a64}{k18}{0}
    \mypsvinvisible{0.152495}{2.09336}{a65}{k18}{0}
    \mypsvinvisible{-0.236085}{2.08559}{a66}{k19}{0}
    \mypsvinvisible{-0.394502}{2.0615}{a67}{k19}{0}
    \mypsvinvisible{-0.767831}{1.95342}{a68}{k20}{0}
    \mypsvinvisible{-0.914615}{1.88915}{a69}{k20}{0}
    \mypsvinvisible{-1.24725}{1.68813}{a70}{k21}{0}
    \mypsvinvisible{-1.3724}{1.58806}{a71}{k21}{0}
    \mypsvinvisible{-1.64167}{1.3078}{a72}{k22}{0}
    \mypsvinvisible{-1.73666}{1.17874}{a73}{k22}{0}
    \mypsvinvisible{-1.92421}{0.838338}{a74}{k23}{0}
    \mypsvinvisible{-1.98256}{0.689101}{a75}{k23}{0}
    \mypsvinvisible{-2.07563}{0.311749}{a76}{k24}{0}
    \mypsvinvisible{-2.09336}{0.152495}{a77}{k24}{0}
    \mypsvinvisible{-2.08559}{-0.236085}{a78}{k25}{0}
    \mypsvinvisible{-2.0615}{-0.394502}{a79}{k25}{0}
    \mypsvinvisible{-1.95342}{-0.767831}{a80}{k26}{0}
    \mypsvinvisible{-1.88915}{-0.914615}{a81}{k26}{0}
    \mypsvinvisible{-1.68813}{-1.24725}{a82}{k27}{0}
    \mypsvinvisible{-1.58806}{-1.3724}{a83}{k27}{0}
    \mypsextra
    \end{pspicture}
    \caption{Radius $6$ neighbourhoods of (a) an objective $k_0 \in K$ and (b) a constraint $i_0 \in I$ in the regularised tree $\myG$, assuming $\Di = 4$ and $\Dk = 3$. The black dots represent agents $v \in V$, the open circles represent objectives $k \in K$, and the boxes represent constraints $i \in I$.}\label{fig:regular}
\end{figure}
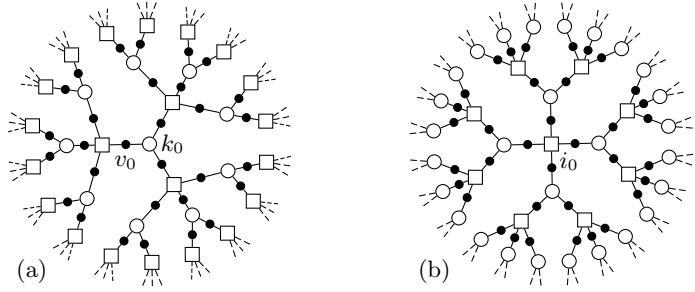

Now repeat these steps and grow virtual trees rooted at the constraints and objectives that had less than $\Di$ or $\Dk$ neighbours. The result is a countably infinite tree where $\mysize{V_i} = \Di$ and $\mysize{V_k} = \Dk$ for all $i \in I$ and $k \in K$. Observe also that from the perspective of a local algorithm it suffices to grow the virtual trees only up to depth $r$ because then the radius $r$ neighbourhood of each original node is indistinguishable from the regularised tree. The resulting topology is illustrated in Fig.~\ref{fig:regular} from the perspective of an original objective $k_0 \in K$ and an original constraint $i_0 \in I$.

\subsection{Properties of regularised trees}

For each $v \in V$ in a regularised tree $\myG$, define
$K(v,\ell) = K \cap B_\myG(v, 4\ell+1)$,
that is, the set of objectives $k$ within distance $4\ell+1$ from $v$. For example, $K(v,1)$ consists of $1$ objective at distance $1$, $\Di-1$ objectives at distance $3$, and $(\Dk-1)(\Di-1)$ objectives at distance $5$; see Fig.~\ref{fig:regular}a. 
In general, we have
\begin{equation}\label{eq:size-Kvl}
    \mysize{K(v,\ell)} = 1 + (\Di - 1)\Dk n(\ell),
\end{equation}
where
\[
    n(\ell) = \textstyle\sum_{j = 0}^{\ell-1} (\Di{-}1)^j (\Dk{-}1)^j .
\]

Let $k \in K$. If $u, v \in V_k$, $u \ne v$, then the objective at distance $1$ from $u$ is the same as the objective at distance $1$ from $v$; therefore $K(u,0) = K(v,0)$. The objectives at distance $3$ from $u$ are at distance $5$ from $v$, and the objectives at distance $5$ from $u$ are at distance $3$ or $5$ from $v$; therefore $K(u,1) = K(v,1)$. By a similar reasoning, we obtain
\begin{equation}\label{eq:KvL-Vk}
    K(u,\ell) = K(v,\ell) \qquad \forall \ \ell \in \mynatural, \  k \in K, \ u, v \in V_k.
\end{equation}

Let us then study a constraint $i \in I$. Define
\[
    K(i,\ell) = \textstyle\bigcap_{v \in V_i} K(v,\ell) = K \cap B_\myG(i, 4\ell) = K \cap B_\myG(i, 4\ell-2) .
\]
For example, $K(i,2)$ consists of $\Di$ objectives at distance $2$ from the constraint $i$, and ${\Di(\Dk-1)(\Di-1)}$ objectives at distance $6$ from the constraint $i$; see Fig.~\ref{fig:regular}b. In general, we have
\begin{equation}\label{eq:size-Kil}
    \mysize{K(i,\ell)} = \Di n(\ell).
\end{equation}
For adjacent $v \in V$ and $i \in I$, we also define $\partial K(v,i,\ell) = K(v,\ell) \setminus K(i,\ell)$. We have by \eqref{eq:size-Kvl} and \eqref{eq:size-Kil}
\begin{equation}\label{eq:size-Kvil}
    \mysize{\partial K(v,i,\ell)} = 1 + (\Di \Dk - \Di - \Dk) \, n(\ell) .
\end{equation}

\subsection{Local approximation on regularised trees}\label{ssec:approx-reg-tree}

It now suffices to meet Lemma~\ref{lem:unfolding} for bipartite max-min LPs in the case when the underlying graph $\myG$ is a countably infinite regularised tree. To this end, let $L \in \mynatural$ be a constant that we choose later; $L$ depends only on $\Di$, $\Dk$ and $\epsilon$.

Each agent $u \in V$ now executes the following algorithm. First, the agent gathers all objectives $k \in K$ within distance $4L+1$, that is, the set $K(u,L)$. Then, for each $k \in K(u,L)$, the agent $u$ gathers the radius $4L+2$ neighbourhood of $k$; let $\myGkL$ be this subgraph. In total, the agent $u$ accumulates information from distance $r = 8L+3$ in the tree; this is the local horizon of the algorithm.

The structure of $\myGkL$ is a tree similar to the one shown in Fig.~\ref{fig:regular}a. The leaf nodes of the tree $\myGkL$ are constraints. For each $k\in K(u,L)$, the agent $u$ forms the constant-size \emph{subproblem} of \eqref{eq:max-min} restricted to the vertices of $\myGkL$ and solves it optimally using a deterministic algorithm; let $x^{\mykL}$ be the solution. Once the agent $u$ has solved the subproblem for every $k \in K(u,L)$, it sets
\begin{align}
    q &= 1 / \big(\Di + \Di (\Di - 1)(\Dk - 1) n(L)\big), \label{eq:approx-q} \\
    x_u &= q \textstyle\sum_{k \in K(u,L)} x^{\mykL}_u. \label{eq:approx-xv}
\end{align}
This completes the description of the algorithm.

We now show that the computed solution $x$ is feasible. Because each $x^{\mykL}$ is a feasible solution, we have
\begin{align}
    \textstyle\sum_{v \in V_{i}} \myaspacing_{iv} x^{\mykL}_v &\le 1 &
    &\forall\ \text{non-leaf $i \in I$ in $\myGkL$},
    \label{eq:sub-feasible-nonleaf}\\
    \myaspacing_{iv} x^{\mykL}_v &\le 1 &
    &\forall\ \text{leaf $i \in I$, $v \in V_i$ in $\myGkL$}.
    \label{eq:sub-feasible-leaf}
\end{align}
Let $i \in I$. For each subproblem $\myGkL$ with $v \in V_i$, $k \in K(i,L)$, the constraint $i$ is a non-leaf vertex; therefore
\begin{equation}
        \sum_{v \in V_i} \ \sum_{k \in K(i,L)} \myaspacing_{iv} x^{\mykL}_v
        \ =\  \sum_{k \in K(i,L)} \ \sum_{v \in V_i} \myaspacing_{iv} x^{\mykL}_v 
        \ \myeqoverset{eq:sub-feasible-nonleaf}{\le}\  \sum_{k \in K(i,L)} 1 
        \ \myeqoverset{eq:size-Kil}{=}\  \Di \, n(L). \label{eq:approx-feasible-nonleaf}
\end{equation}
For each subproblem $\myGkL$ with $v \in V_i$, $k \in \partial K(v,i,L)$, the constraint $i$ is a leaf vertex; therefore
\begin{align}
        \sum_{v \in V_i} \ \sum_{k \in \partial K(v,i,L)} \myaspacing_{iv} x^{\mykL}_v
        &\ \myeqoverset{eq:sub-feasible-leaf}{\le}\  \sum_{v \in V_i} \ \sum_{k \in \partial K(v,i,L)} 1 \nonumber\\
        &\ \myeqoverset{eq:size-Kvil}{=}\  \Di \, (1 + (\Di \Dk - \Di - \Dk) \, n(L)). \label{eq:approx-feasible-leaf}
\end{align}
Combining \eqref{eq:approx-feasible-nonleaf} and \eqref{eq:approx-feasible-leaf}, we can show that the constraint $i$ is satisfied:
\begin{align*}
    \sum_{v \in V_i} a_{iv} x_v
        &\ \myeqoverset{eq:approx-xv}{=}\  q \sum_{v \in V_i} a_{iv} \sum_{k \in K(v,L)} x^{\mykL}_v \\
        &\ =\  q \bigg( \sum_{v \in V_i} \ \sum_{k \in K(i,L)} \myaspacing_{iv} x^{\mykL}_v \bigg)
            + q \bigg( \sum_{v \in V_i} \ \sum_{k \in \partial K(v,i,L)} \myaspacing_{iv} x^{\mykL}_v \bigg) \\
        &\ \le\ q \Di n(L) \ + \  q \Di (1 + (\Di \Dk - \Di - \Dk) n(L))
        \ \myeqoverset{eq:approx-q}{=}\ 1.
\end{align*}

Next we establish a lower bound on the performance of the algorithm. To this end, consider an arbitrary feasible solution $x'$ of the unrestricted problem \eqref{eq:max-min} with utility at least $\myutil'$. This feasible solution is also a feasible solution of each finite subproblem restricted to $\myGkL$; therefore
\begin{align}
    \textstyle\sum_{v \in V_h} \mycspacing_{hv} x^{\mykL}_v &\ge \myutil' &
    &\forall\ \text{$h \in K$ in $\myGkL$}.
    \label{eq:sub-optimal}
\end{align}
Define
\begin{equation}\label{eq:approx-alpha}
    \alpha \ =\  \frac{1}{q (1 + (\Di - 1)\Dk n(L))} \ \myeqoverset{eq:approx-q}{=}\  \Di \left(1 - \frac{1}{\Dk + 1 / ((\Di - 1) n(L))} \right) .
\end{equation}
Consider an arbitrary $k \in K$ and $u \in V_k$. We have
\begin{align*}
    \sum_{v \in V_k} c_{kv} x_v
        &\ =\  q \sum_{v \in V_k} c_{kv} \sum_{h \in K(v,L)} x^{\myhL}_v
        \ \myeqoverset{eq:KvL-Vk}{=}\  q \sum_{h \in K(u,L)} \ \sum_{v \in V_k} \mycspacing_{kv} x^{\myhL}_v \nonumber\\
        &\ \myeqoverset{eq:sub-optimal}{\ge}\  q \sum_{h \in K(u,L)} \myutil'
        \ \myeqoverset{eq:size-Kvl}{\ge}\  q (1 + (\Di - 1)\Dk n(L)) \, \myutil'
        \ \myeqoverset{eq:approx-alpha}{=}\ \myutil' / \alpha .
\end{align*}
For a sufficiently large $L$, we meet Lemma~\ref{lem:unfolding} with $\alpha < {\Di (1 - 1/\Dk) + \epsilon}$. This completes the proof of Theorem~\ref{thm:approx}. For a concrete example, see Appendix~\ref{app:approx}.

\section{Inapproximability results}\label{sec:inapprox}

We proceed to prove Theorems \ref{thm:inapprox} and \ref{thm:inapprox-bounded}. Let $r=4,8,\ldots$, $\mylength \in \mynatural$, $\myPI \in \myposinteger$, and $\myPK \in \myposinteger$ be constants whose values we choose later. Let $\myQ = (\myIB \cup \myKB, \myEB)$ be a bipartite graph where the degree of each $i \in \myIB$ is $\myPI$, the degree of each $k \in \myKB$ is $\myPK$, and there is no cycle of length less than $\mygirth = 2(4\mylength+2+r)+1$. Such graphs exist for all values of the parameters; a simple existence proof can be devised by slightly modifying the proof of a theorem of Hoory \cite[Theorem~A.2]{hoory02graphs}; see Appendix~\ref{app:high-girth}.

\subsection{The instance $\myinst$}\label{ssec:inst-S}

Given the graph $\myQ = (\myIB \cup \myKB, \myEB)$, we construct an instance of the max-min LP problem, $\myinst$. The underlying communication graph $\myG = (V \cup I \cup K, E)$ is constructed as shown in the following figure.
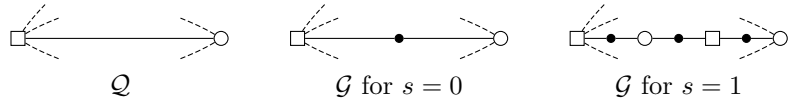
\begin{center}
    \begin{pspicture}(-0.7,0.3)(3.3,1.5)%
        \mypsgrid
        \rput[cB](1.5,0.3){$\myQ$}
        \mypsI
        \mypsi{0.0}{1.0}{i1}
        \mypsK
        \mypsk{3.0}{1.0}{k1}
        \pnode(0.6, 1.3){ix1}
        \pnode(0.4, 1.5){ix2}
        \pnode(0.6, 0.7){ix3}
        \pnode(2.4, 1.3){kx1}
        \pnode(2.4, 0.7){kx2}
        \mypsgenericedges
        \ncarc[arcangle=0]{-}{i1}{k1}
        \mypsgenericedgesdashed
        \ncarc[arcangle=5]{-}{i1}{ix1}
        \ncarc[arcangle=5]{-}{i1}{ix2}
        \ncarc[arcangle=-5]{-}{i1}{ix3}
        \ncarc[arcangle=-5]{-}{k1}{kx1}
        \ncarc[arcangle=5]{-}{k1}{kx2}
    \end{pspicture}
    \begin{pspicture}(-0.7,0.3)(3.3,1.5)%
        \mypsgrid
        \rput[cB](1.5,0.3){$\myG$ for $\mylength = 0$}
        \mypsI
        \mypsi{0.0}{1.0}{i1}
        \mypsK
        \mypsk{3.0}{1.0}{k1}
        \mypsV
        \mypsv{1.5}{1.0}{v1}{i1}{k1}{0}{0}
        \mypsVinvisible
        \mypsvinvisible{0.6}{1.3}{vinv1}{i1}{-5}
        \mypsvinvisible{0.4}{1.5}{vinv2}{i1}{-5}
        \mypsvinvisible{0.6}{0.7}{vinv3}{i1}{5}
        \mypsvinvisible{2.4}{1.3}{vinv4}{k1}{5}
        \mypsvinvisible{2.4}{0.7}{vinv4}{k1}{-5}
    \end{pspicture}
    \begin{pspicture}(-0.7,0.3)(3.3,1.5)%
        \mypsgrid
        \rput[cB](1.5,0.3){$\myG$ for $\mylength = 1$}
        \mypsI
        \mypsi{0.0}{1.0}{i1}
        \mypsi{2.0}{1.0}{inew1}
        \mypsK
        \mypsk{3.0}{1.0}{k1}
        \mypsk{1.0}{1.0}{knew1}
        \mypsV
        \mypsv{0.5}{1.0}{v1a}{i1}{knew1}{0}{0}
        \mypsv{1.5}{1.0}{v1b}{knew1}{inew1}{0}{0}
        \mypsv{2.5}{1.0}{v1c}{inew1}{k1}{0}{0}
        \mypsVinvisible
        \mypsvinvisible{0.6}{1.3}{vinv1}{i1}{-5}
        \mypsvinvisible{0.4}{1.5}{vinv2}{i1}{-5}
        \mypsvinvisible{0.6}{0.7}{vinv3}{i1}{5}
        \mypsvinvisible{2.4}{1.3}{vinv4}{k1}{5}
        \mypsvinvisible{2.4}{0.7}{vinv4}{k1}{-5}
    \end{pspicture}
\end{center}
Each edge $e = \{i,k\} \in \myEB$ is replaced by a path of length $4 \mylength + 2$: the path begins with the constraint $i \in \myIB$; then there are $\mylength$ segments of agent--objective--agent--constraint; and finally there is an agent and the objective $k \in \myKB$. There are no other edges or vertices in $\myG$. For example, in the case of $\mylength = 0$, $\myPI = 4$, $\myPK = 3$, and sufficiently large $\mygirth$, the graph $\myG$ looks \emph{locally} similar to the trees in Fig.~\ref{fig:regular}, even though there may be long cycles.

The coefficients of the instance $\myinst$ are chosen as follows. For each objective $k \in \myKB$, we set $c_{kv} = 1$ for all $v \in V_k$. For each objective $k \in K \setminus \myKB$, we set $c_{kv} = \myPK - 1$ for all $v \in V_k$. For each constraint $i \in I$, we set $a_{iv} = 1$. Observe that $\myinst$ is a bipartite max-min LP; furthermore, in the case $\mylength = 0$, this is a 0/1 max-min LP. We can choose the port numbering in $\myG$ in an arbitrary manner, and we can assign unique node identifiers to the vertices of $\myG$ as well.

Consider a feasible solution $x$ of $\myinst$, with utility $\myutil$. We proceed to derive an upper bound for~$\myutil$. For each $j = 0, 1, \ldots, 2\mylength$, let $V(j)$ consist of agents $v \in V$ such that the distance to the nearest constraint $i \in \myIB$ is $2j + 1$. That is, $V(0)$ consists of the agents adjacent to an $i \in \myIB$ and $V(2\mylength)$ consists of the agents adjacent to a $k \in \myKB$. Let $m = \mysize{\myEB}$; we observe that $\mysize{V(j)} = m$ for each $j$.

Let $X(j) = \sum_{v \in V(j)} x_v / m$. From the constraints $i \in \myIB$ we obtain
\[
    X(0)
    \,= \sum_{v \in V(0)} x_v / m
    \,=\, \sum_{i \in \myIB} \sum_{v \in V_i} a_{iv} x_v / m
    \,\le\, \sum_{i \in \myIB} 1 / m
    \,=\, \mysize{\myIB} / m
    \,=\, 1 / \myPI .
\]
Similarly, from the objectives $k \in \myKB$ we obtain
$X(2\mylength) \ge \myutil \mysize{\myKB} / m = \myutil / \myPK $.

From the objectives $k \in K \setminus \myKB$, taking into account our choice of the coefficients $c_{kv}$, we obtain the inequality
$X(2t) + X(2t + 1) \ge \myutil / (\myPK - 1)$ for $t=0,1,\ldots,s-1$.
From the constraints $i \in I \setminus \myIB$, we obtain the inequality
$X(2t + 1) + X(2t + 2) \le 1$ for $t=0,1,\ldots,s-1$.
Combining inequalities, we have
\[
    \begin{split}
        \myutil / \myPK - 1 / \myPI
        &\,\le\, X(2\mylength) - X(0) \\
        &\,=\, \sum_{t = 0}^{\mylength-1} \Big(\big(X(2t + 1) + X(2t + 2)\big) - \big(X(2t) + X(2t + 1)\big)\Big) \\
        &\,\le\, \mylength \cdot \big(1 - \myutil / (\myPK - 1)\big) ,
    \end{split}
\]
which implies
\begin{equation}\label{eq:s-util-max}
    \myutil \le \frac{\myPK}{\myPI} \cdot \frac{\myPK - 1 + \myPK \myPI \mylength - \myPI \mylength}{\myPK - 1 + \myPK \mylength} .
\end{equation}

\subsection{The instance $\myinst_k$}\label{ssec:inst-Sk}

Let $k \in \myKB$. We construct another instance of the max-min LP problem, $\myinst_k$. The communication graph of $\myinst_k$ is the subgraph $\myG_k$ of $\myG$ induced by $B_\myG(k, 4\mylength+2+r)$. By the choice of $\mygirth$, there is no cycle in $\myG_k$. As $r$ is a multiple of $4$, the leaves of the tree $\myG_k$ are constraints. For example, in the case of $\mylength=0$, $\myPI = 4$, $\myPK = 3$, and $r = 4$, the graph $\myG_k$ is isomorphic to the tree of Fig.~\ref{fig:regular}a. The coefficients, port numbers and node identifiers are chosen in $\myG_k$ exactly as in $\myG$. The optimum of $\myinst_k$ is greater than $\myPK-1$ (see Appendix~\ref{app:inst-Sk-util}).

\subsection{Proof of Theorem~\ref{thm:inapprox}}

Let $\Di \ge 2$ and $\Dk \ge 2$. Assume that $\myA$ is a local approximation algorithm with the approximation ratio $\alpha$. Set $\myPI = \Di$, $\myPK = \Dk$ and $\mylength = 0$. Let $r$ be the local horizon of the algorithm, rounded up to a multiple of $4$. Construct the instance $\myinst$ as described in Sect.~\ref{ssec:inst-S}; it is a 0/1 bipartite max-min LP, and it satisfies the degree bounds $\Di$ and $\Dk$. Apply the algorithm $\myA$ to $\myinst$. The algorithm produces a feasible solution $x$. By \eqref{eq:s-util-max} there is a constraint $k$ such that $\sum_{v \in V_k} x_v \le \Dk / \Di$.

Now construct $\myinst_k$ as described in Sect.~\ref{ssec:inst-Sk}; this is another 0/1 bipartite max-min LP. Apply $\myA$ to $\myinst_k$. The algorithm produces a feasible solution $x'$. The radius $r$ neighbourhoods of the agents $v \in V_k$ are identical in $\myinst$ and $\myinst_k$; therefore the algorithm must make the same decisions for them, and we have $\sum_{v \in V_k} x'_v \le \Dk / \Di$. But there is a feasible solution of $\myinst_k$ with utility greater than $\Dk-1$ (see Appendix~\ref{app:inst-Sk-util}); therefore the approximation ratio of $\myA$ is $\alpha > {(\Dk-1)} /\allowbreak {(\Dk / \Di)}$. This completes the proof of Theorem~\ref{thm:inapprox}.

\subsection{Proof of Theorem~\ref{thm:inapprox-bounded}}

Let $\Di \ge 3$, $\Dk \ge 3$, and $0 < \delta < 1/10$. Assume that $\myA$ is a local approximation algorithm with the approximation ratio $\alpha$. Set $\myPI = 3$, $\myPK = 3$, and $\mylength = {\myceil{4 / (7\delta) - 1/2}}$. Let $r$ be the local horizon of the algorithm, rounded up to a multiple of $4$.

Again, construct the instance $\myinst$. The relative growth of $\myG$ is at most $1 + {2^j / ((2^j-1) (2 \mylength + 1))}$ beyond radius $R = {j (4 \mylength + 2)}$; indeed, each set of $2^j$ new agents can be accounted for $1 + 2 + \dotsb + 2^{j-1}=2^j-1$ chains with $2\mylength + 1$ agents each. Choosing $j = 3$, the relative growth of $\myG$ is at most $1 + \delta$ beyond radius~$R$.

Apply $\myA$ to $\myinst$. By \eqref{eq:s-util-max} we know that there exists an objective $h$ such that $\sum_{v \in V_{h}} x_v \le 2 - 2 / (3 \mylength + 2)$. Choose a $k \in \myKB$ nearest to $h$. Construct $\myinst_k$ and apply $\myA$ to $\myinst_k$. The local neighbourhoods of the agents $v \in V_{h}$ are identical in $\myinst$ and $\myinst_k$. We know that $\myinst_k$ has a feasible solution with utility greater than $2$ (see Appendix~\ref{app:inst-Sk-util}). Using the assumption $\delta < 1/10$, we obtain
\[
    \alpha > \frac{2}{2-2/(3s+2)} = 1 + \frac{1}{3 \mylength + 1} \ge 1 + \frac{1}{3 (4 / (7\delta) + 1/2) + 1} > 1 + \frac{\delta}{2} .
\]

\subsubsection*{Acknowledgements.}

This research was supported in part by the Academy of Finland, Grants 116547 and 117499, and by Helsinki Graduate School in Computer Science and Engineering (Hecse).

\providecommand{\noopsort}[1]{}

\clearpage
\appendix

\section{Appendix}

\subsection{Unfolding in graph theory and topology}\label{app:terminology}

We briefly summarise the graph theoretic and topological background related to the unfolding $(\myT, f)$ of $\myH$ as defined in Sect.~\ref{sec:unfolding}.

From a graph theoretic perspective, using the terminology of Godsil and Royle~\cite[\S6.8]{godsil04algebraic}, surjection $f$ is a homomorphism from $\myT$ to $\myH$. Moreover, it is a \emph{local isomorphism}: the neighbours of $\myt v \in \myt V$ are in one-to-one correspondence with the neighbours of $f(\myt v) \in V$. A surjective local isomorphism $f$ is a \emph{covering map} and $(\myT, f)$ is a \emph{covering graph} of $\myH$.

Covering maps in graph theory can be interpreted as a special case of covering maps in topology: $\myT$ is a \emph{covering space} of $\myH$ and $f$ is, again, a covering map. See, e.g., Hocking and Young~\cite[\S4.8]{hocking61topology} or Munkres~\cite[\S53]{munkres00topology}.

In topology, a simply connected covering space is called a \emph{universal covering space} \cite[\S4.8]{hocking61topology}, \cite[\S80]{munkres00topology}. An analogous graph-theoretic concept is a tree: unfolding $\myT$ of $\myH$ is equal to the \emph{universal covering} $\mathcal{U}(\myH)$ of $\myH$ as defined by Angluin~\cite{angluin80local}.

Unfortunately, the term ``covering'' is likely to cause confusion in the context of graphs. The term ``lift'' has been used for a covering graph~\cite{hoory02graphs,amit01random}. We have borrowed the term ``unfolding'' from the field of model checking; see, e.g., Esparza and Heljanko~\cite{esparza00new}.

\subsection{Infinite max-min LPs}\label{app:infinite}

Unfolding (Sect.~\ref{sec:unfolding}) and regularisation (Sect.~\ref{sec:approx}) in general require us to consider max-min LPs where the underlying graph $\myG$ is countably infinite. Observe that $\myG$ is always a bounded-degree graph, however. This allows us to circumvent essentially all of the technicalities otherwise encountered with infinite problem instances; cf. Anderson and Nash~\cite{anderson87linear}.

For the purposes of this work, it suffices to define that $x$ is a \emph{feasible solution with utility at least} $\myutil$ if $(x,\myutil)$ satisfies
\begin{equation}
    \begin{aligned}
        \textstyle\sum_{v \in V_k} c_{kv} x_v \,&\ge\, \myutil & \quad\forall\, &k \in K, \\
        \textstyle\sum_{v \in V_i} a_{iv} x_v \,&\le\, 1 & \quad\forall\, &i \in I, \\
        x_v \,&\ge\, 0 & \forall\, &v \in V .
    \end{aligned}\label{eq:max-min-infinite}
\end{equation}
Each of the sums in \eqref{eq:max-min-infinite} is finite.

Observe that this definition is compatible with the finite max-min LP defined in Sect.~\ref{ssec:max-min}. Namely, if $\myoptutil$ is the optimum of a finite max-min LP, then there exists a feasible solution $\myopt{x}$ with utility at least $\myoptutil$.

\subsection{Proof of Lemma~\ref{lem:unfolding}}\label{app:unfolding}

Assume that an arbitrary finite max-min LP from the family under consideration is given as input. Let $\myG = (V \cup I \cup K, E)$ be the underlying communication graph. Unfold $\myG$ to obtain a (possibly infinite) tree $\myT = (\myt V \cup \myt I \cup \myt K, \myt E)$ with a labelling $f$. Extend this to an unfolding of the max-min LP by associating a variable $x_{\myt v}$ with each agent $\myt v \in \myt V$, the coefficient $a_{\myti \myt v}=a_{f(\myti), f(\myt v)}$ for each edge $\{\myti, \myt v\} \in \myt E$, $\myti \in \myt I$, $\myt v \in \myt V$, and the coefficient $c_{\mytk \myt v}=c_{f(\mytk), f(\myt v)}$ for each edge $\{\mytk, \myt v\} \in \myt E$, $\mytk \in \myt K$, $\myt v \in \myt V$. Furthermore, assume an arbitrary port numbering for the edges incident to each of the nodes in $\myG$, and extend this to a port numbering for the edges incident to each of the nodes in $\myT$ so that the port numbers at the ends of each edge $\{\myt u,\myt v\}\in \myt E$ are identical to the port numbers at the ends of $\{f(\myt u),f(\myt v)\}$. 

Let $\myopt x$ be an optimal solution of the original instance, with utility $\myoptutil$. Set $x_{\myt v} = \myopt{x}_{f(\myt v)}$ to obtain a solution of the unfolding. This is a feasible solution because the variables of the agents adjacent to a constraint $\myti$ in the unfolding have the same values as the variables of the agents adjacent to the constraint $f(\myti)$ in the original instance. By similar reasoning, we can show that this is a feasible solution with utility at least $\myoptutil$.

Construct the local algorithm $\myA$ using the assumed algorithm $\myt \myA$ as follows. Each node $v \in V$ simply behaves as if it was a node $\myt v \in f^{-1}(v)$ in the unfolding $\myT$ and simulates $\myt \myA$ for $\myt v$ in $\myT$. By assumption, the solution $x$ computed by $\myt \myA$ in the unfolding has to satisfy $\sum_{\myt v \in V_{\mytk}} c_{\mytk \myt v} x_{\myt v} \ge \myoptutil/\alpha$ for every $\mytk \in \myt K$ and $\sum_{\myt v \in V_{\myti}} a_{\myti \myt v} x_{\myt v} \le 1$ for every $\myti \in \myt I$. Furthermore, if $f(\myt u) = f(\myt v)$ for $\myt u, \myt v \in \myt V$, then the neighbourhoods of $\myt u$ and $\myt v$ contain precisely the same information (including the port numbering), so the deterministic $\myt\myA$ must output the same value $x_{\myt u} = x_{\myt v}$. Giving the output $x_v = x_{\myt v}$ for any $\myt v \in f^{-1}(v)$ therefore yields a feasible, $\alpha$-approximate solution to the original instance. This completes the proof.

We observe that Lemma~\ref{lem:unfolding} generalises beyond max-min LPs; we did not exploit the linearity of the constraints and the objectives.

\subsection{The approximation algorithm in practice}\label{app:approx}

In this section, we give a simple example that illustrates the behaviour of the approximation algorithm presented in Sect.~\ref{ssec:approx-reg-tree}. Consider the case of $\Di = 4$, $\Dk = 3$ and $L = 1$. For each $k \in K$, we construct and solve a subproblem; the structure of the subproblem is illustrated in Fig.~\ref{fig:regular}a. Then we simply sum up the optimal solutions of each subproblem. For any $v \in V$, the variable $x_v$ is involved in exactly $\mysize{K(v,L)} = 10$ subproblems.

First, consider an objective $k \in K$. The boundary of a subproblem always lies at a constraint, never at an objective. Therefore the objective $k$ and all its adjacent agents $v \in V_k$ are involved in $10$ subproblems. We satisfy the objective exactly $10$ times, each time at least as well as in the global optimum.

Second, consider a constraint $i \in I$. The constraint may lie in the middle of a subproblem or at the boundary of a subproblem. The former happens in this case $\mysize{K(i,L)} = 4$ times; the latter happens $\mysize{V_i} \cdot \mysize{\partial K(v,i,L)} = 24$ times. In total, we use up the capacity available at the constraint $i$ exactly $28$ times. See Fig.~\ref{fig:regular}b for an illustration; there are $28$ objectives within distance $6$ from the constraint $i_0 \in I$.

Finally, we scale down the solution by factor $q = 1/28$. This way we obtain a solution which is feasible and within factor $\alpha = 2.8$ of optimum. This is close to the lower bound $\alpha > 2.66$ from Theorem~\ref{thm:inapprox}.

\subsection{Bipartite high girth graphs}\label{app:high-girth}

We say that a bipartite graph $\myG = (V \cup U, E)$ is $(a,b)$-regular if the degree of each node in $V$ is $a$ and the degree of each node in $U$ is $b$. Here we sketch a proof which shows that for any positive integers $a$, $b$ and $\mygirth$, there is a $(a,b)$-regular bipartite graph $\myG = (V \cup U, E)$ which has no cycle of length less than $\mygirth$. We slightly adapt a proof of a similar result for $d$-regular graphs~\cite[Theorem~A.2]{hoory02graphs} to our needs. We proceed by induction on $\mygirth$, for $\mygirth = 4, 6, 8, \dotsc$.

First consider the basis $\mygirth = 4$. We can simply choose the complete bipartite graph $K_{b,a}$ as a $(a,b)$-regular graph~$\myG$.

Next consider $\mygirth \ge 6$. Let $\myG = (V \cup U, E)$ be an $(a,b)$-regular bipartite graph where the length of the shortest cycle is $c \ge \mygirth - 2$. Let $S \subseteq E$. We construct a graph $\myG_S = (V_S \cup U_S, E_S)$ as follows:
\begin{align*}
    V_S &= \{0,1\} \times V, \\
    U_S &= \{0,1\} \times U, \\
    E_S &= \{ \{ (0,v), (0,u) \}, \{ (1,v), (1,u) \} : \{v,u\} \in S \} \\
        &\quad \cup \{ \{ (0,v), (1,u) \}, \{ (1,v), (0,u) \} : \{v,u\} \in E \setminus S \}.
\end{align*}
The graph $\myG_S$ is an $(a,b)$-regular bipartite graph (actually, it is a covering graph of $\myG$; see Appendix~\ref{app:terminology}). Furthermore, $\myG_S$ has no cycle of length less than $c$. We proceed to show that there exists a subset $S$ such that the number of cycles of length exactly $c$ in $\myG_S$ is strictly less than the number of cycles of length $c$ in $\myG$. Then by a repeated application of the same construction, we can conclude that there exists a graph which is an $(a,b)$-regular bipartite graph and which has no cycle of length $c$; that is, its girth is at least $\mygirth$.

We use the probabilistic method to show that the number of cycles of length $c$ decreases for some $S \subseteq E$. For each $e \in E$, toss an independent and unbiased coin to determine whether $e \in S$. For each cycle $C \subseteq E$ of length $c$ in $\myG$, we have in $\myG_S$ either two cycles of length $c$ or one cycle of length $2c$, depending on the parity of $\mysize{C \cap S}$.

The expected number of cycles of length $c$ in $\myG_S$ is therefore equal to the number of cycles of length $c$ in $\myG$. The choice $S = E$ doubles the number of such cycles; therefore some other choice necessarily decreases the number of such cycles. This completes the proof.

\subsection{A feasible solution of the instance $\myinst_k$}\label{app:inst-Sk-util}

Consider the instance $\myinst_k$ constructed in Sect.~\ref{ssec:inst-Sk}. We construct a solution $x$ as follows. Let $D = \max\,\{ \myPI, \myPK + 1 \}$. If the distance between the agent $v$ and the objective $k$ in $\myG_k$ is $4j+1$ for some $j$, set $x_v = 1 - 1/D^{2j+1}$. If the distance is $4j+3$, set $x_v = 1/D^{2j+2}$.

This is a feasible solution. Feasibility is clear for each leaf constraint $i \in I$. Then consider a non-leaf constraint $i \in I$. They have at most $\myPI$ neighbours, and the distance between $k$ and $i$ is $4j + 2$ for some $j$. Thus
\[
    \sum_{v \in V_i} a_{iv} x_v \le 1 - 1/D^{2j+1} + (\myPI-1)/D^{2j+2} < 1.
\]

Let $\myutil_k$ be the utility of this solution. We show that $\myutil_k > \myPK-1$. First, consider the objective $k$. We have
\[
    \sum_{v \in V_k} c_{kv} x_v = \myPK (1 - 1/D ) > \myPK - 1.
\]
Second, consider an objective $h \in \myKB \setminus \{k\}$. It has $\myPK$ neighbours and the distance between $h$ and $k$ is $4j$ for some $j$. Thus
\[
    \sum_{v \in V_{h}} c_{hv} x_v = 1/D^{2j} + (\myPK-1) (1 - 1/D^{2j+1}) > \myPK - 1.
\]
Finally, consider an objective $h \in K \setminus \myKB$. It has $2$ neighbours and the distance between $h$ and $k$ is $4j$ for some $j$; the coefficients are $c_{hv} = \myPK-1$. Thus
\[
    \sum_{v \in V_{h}} c_{hv} x_v = (\myPK-1) (1/D^{2j} + 1 - 1/D^{2j+1}) > \myPK - 1 .
\]

\end{document}